\documentclass[twocolumn,floatfix,prl,aps,showpacs,superscriptaddress]{revtex4}
\usepackage{epsfig,graphicx}
\usepackage{flafter}
\usepackage{amsmath}
\usepackage{xcolor}
\usepackage{bm}
\usepackage{amssymb}
\usepackage{epstopdf}
\usepackage{amstext}
\usepackage{amsthm}
\usepackage{latexsym}
\usepackage{multirow}
\usepackage{mathtools}
\usepackage{bm}
\usepackage{bbm}
    \usepackage{amsmath}
    \usepackage{amsthm}
    \usepackage{amsfonts}
    \usepackage{braket}
\usepackage{soul}
\usepackage{comment}

\begin{document}

\title{Creation and robustness of quantized vortices in a dipolar supersolid when crossing the superfluid-to-supersolid transition} 

\author{Marija \v{S}indik}
\affiliation{INO-CNR BEC Center and Dipartimento di Fisica, Universit\`a degli Studi di Trento, 38123 Povo, Italy}
\affiliation{Institute of Physics Belgrade, University of Belgrade, Serbia}

\author{Alessio Recati\footnote{Corresponding Author: alessio.recati@ino.it}}
\affiliation{INO-CNR BEC Center and Dipartimento di Fisica, Universit\`a degli Studi di Trento, 38123 Povo, Italy}
\affiliation{Trento  Institute  for  Fundamental  Physics  and  Applications,  INFN,  38123 Trento,  Italy}

\author{Santo Maria Roccuzzo}
\affiliation{Kirchhoff-Institut fur Physik, Ruprecht-Karls-Universitat Heidelberg, 69120 Heidelberg, Germany}
\author{Luis Santos\footnote{Corresponding Author: santos@itp.uni-hannover.de}}
\affiliation{Institut f\"ur Theoretische Physik, Leibniz Universit\"at Hannover, Germany}
\author{Sandro Stringari}
\affiliation{INO-CNR BEC Center and Dipartimento di Fisica, Universit\`a degli Studi di Trento, 38123 Povo, Italy}
\affiliation{Trento  Institute  for  Fundamental  Physics  and  Applications,  INFN,  38123 Trento,  Italy}

\date{\today}

\begin{abstract}
Experiments on dipolar Bose-Einstein condensates have recently reported the observation of supersolidity. 
Although quantized vortices constitute a key  probe of superfluidity, their observability in dipolar supersolids is largely prevented by 
the strong density depletion caused by the formation of droplets. We present a novel approach 
to the nucleation of vortices and their observation, based on the quenching of the s-wave scattering length across the superfluid-supersolid transition. 
Starting from a slowly rotating, vortex-free, configuration in the superfluid phase, we predict vortex nucleation as the system enters the supersolid phase, due to 
the strong reduction of the critical angular velocity in the supersolid. Once a vortex is created, we show that it is robustly preserved when the condensate is brought back 
to the superfluid phase, where it may be readily observed.
\end{abstract}

\maketitle



Quantized vortices constitute a key hallmark of superfluidity~\cite{BecBook2016}. They are topological defects of the order parameter, 
and therefore robust with respect to perturbations in the $U(1)$ broken symmetry phase. 
Ultra-cold gases are an ideal platform for the study of vortices. In these gases, vortices are typically created either by stirring the cloud with a laser, or by rotating a slightly deformed trap.
Vortices are detected, after a condensate expansion, by the observation of the density holes corresponding to the vortex cores. Vortices have been observed both in Bose-Einstein condensates~(BECs)~\cite{Cornell1999,dalibard2000,Abo-Shaeer476} and in superfluid spin-$1/2$ Fermi gases~\cite{Zwierlein05}. The angular momentum and its quantization in presence of a vortex can be inferred by exploiting the lift in the degeneracy of quadrupole-mode frequencies due to broken time-reversal symmetry~\cite{zambelli98}, as observed in condensates~\cite{chevy2000,Cornell2001}.


Supersolids constitute a particularly intriguing phase in which superfluidity coexists with a modulated density~\cite{Boninsegni2012}. Supersolidity 
has attracted in the last few years a major attention in ultra-cold gases. Experiments on BECs in optical cavities have revealed supersolid-like properties~\cite{Leonard2017}. 
Condensates with an imposed one-dimensional spin-orbit coupling have been shown to present a supersolid stripe phase~\cite{Li2017,Geier2021}. 
Recent experiments on BECs of magnetic atoms have revealed the creation of supersolids of ultra-dilute droplets maintained by the 
interplay between attractive mean-field interactions and the effective repulsion induced by quantum fluctuations~\cite{Boettcher2020}. 
Dipolar supersolids have attracted a quickly growing interest, and successful experiments in droplet arrays have studied the phase coherence~\cite{F1,S1,I1,Itzhoeffer2021},
 the appearance of Goldstone modes in the excitation spectrum~\cite{F2,S2,I2}, and the peculiar dynamics related to scissors modes~\cite{PfauScissor,Tanzi2021_SF,Norcia2021_scissors,Roccuzzo2021_inertia}. 
Very recently two dimensional supersolid configurations have been also realised \cite{Norcia2021_2D,Bland2021_2D}.



Recent theoretical works have investigated quantum vortices in dipolar supersolids~\cite{gallemi20,ancilotto21}. Quantum vortices in a supersolid were first discussed in Ref.~\cite{Pomeau94} in the context of an hypothetical supersolid phase of Helium. There it was shown, in the context of a mean-field Gross-Pitaevskii formalism employing a repulsive soft-core interaction, that vortices may be nucleated in the supersolid by an obstacle. It was suggested as well that vortices could be robust when crossing back and forth the superfluid-to-supersolid transition. A peculiar feature pointed out in Ref.~\cite{gallemi20} in the case of supersolid dipolar gases is that vortices are, both energetically and dynamically, more favored in the supersolid phase than in the superfluid one. The low-density regions surrounding the droplets of the supersolid phase help in reducing the energetic barrier for a vortex to enter the system, and in pinning the vortices in the interstitials between droplets~\cite{gallemi20}. Even a very slow rotation of the trapping potential can then trigger the dynamical instability that drives vortex nucleation~\cite{gallemi20}. However, the direct detection of vortices formed in the interstitials is largely inhibited because, even in the absence of vortices, this region is characterized by a very low density.



In this Letter, we first explore in detail the robustness of vortices in dipolar BECs when crossing the superfluid-supersolid transition, showing that the conservation of angular momentum results in a peculiar dynamic behavior since the value of the angular momentum per particle associated to a vortex is markedly different in the superfluid and in the supersolid phase. Using the difference in the vortex properties in both phases, we propose a novel dynamic protocol based on the quench of a slowly rotating dipolar condensate from the superfluid into the supersolid phase. A vortex is nucleated in the supersolid due to the strongly reduced critical angular velocity, and a subsequent quench back allows for straightforward vortex imaging in the superfluid phase. Our protocol could provide not only the experimental proof of vortex nucleation in a dipolar supersolid, but also allows for directly probing the modified vortex properties in that phase~\cite{gallemi20}, as, e.g., the reduction of the critical angular velocity for vortex nucleation. It has the advantage of avoiding the nucleation of vortices starting from the equilibrium configuration in the supersolid phase, whose implementation is notoriously more difficult due to three-body collisions.



\paragraph{Model.--} We consider a BEC of atoms with mass $m$ and magnetic 
dipole moment $\mu$ aligned along the $z$ axis, trapped in a harmonic potential of the form  
$V_{ext}({\bf r})=m\omega_\perp^2\left[ (1-\varepsilon)x^2+(1+\varepsilon) y^2+\lambda^2 z^2\right]/2$. 

At zero temperature the physics of the system is well described by 
 the extended Gross-Pitaevskii equation (eGPE)~\cite{Pelster,Wachtler2016}:
\begin{align}
    &i\hbar\frac{\partial \Psi({\bf r},t)}{\partial t} = 
     \Bigl[-\frac{\hbar^2\nabla^2}{2m}+V_{ext}({\bf r}) + g|\Psi({\bf r},t)|^2 \nonumber \\
    & + \int d{\bf r}' V_{dd}({\bf r}-{\bf r}')|\Psi({\bf r}',t)|^2
 + \gamma|\Psi({\bf r},t)|^3 \Bigr] \Psi({\bf r},t),
 \label{eGP}
\end{align}
where $g=4\pi\hbar^2a/m>0$ is the coupling constant fixed by the s-wave scattering length $a$, and 
$V_{dd}({\bf r}) =\frac{\mu_0\mu^2}{4\pi}\frac{1-3\cos^2\theta}{|{\bf r}|^3}$ is the dipole-dipole interaction, with $\theta$ 
the angle between ${\bf r}$ and the $z$ axis. The last term in Eq.~\eqref{eGP} is given by the repulsive Lee-Huang-Yang (LHY) correction 
induced by quantum fluctuations, with 
\begin{equation}
    \gamma=\frac{32 ga^{3/2}}{3\sqrt{\pi}}{\mathrm{ Re}}\left[\int_0^1  du
    [1+\epsilon_{dd}(3u^2-1)]^{5/2}\right],
\end{equation}
where $\epsilon_{dd} =\mu_0\mu^2/3g$ characterizes the relative strength of the dipolar interaction with respect to the contact one.
The eGPE has been systematically employed in the last few years to investigate quantum droplets and supersolidity in dipolar BECs~\cite{Boettcher2020}.  

For small-enough $\epsilon_{dd}$, 
the system behaves as a standard condensate~(superfluid phase). 
By decreasing the scattering length, and hence increasing the value of $\epsilon_{dd}$, the role of the attractive part of the dipolar force becomes more important, and the LHY term starts playing a crucial role in determining the equilibrium solution. The LHY term ensures the stability of the system against collapse and eventually favors the formation of a periodic structure, which can be regarded as a series of dense droplets connected by a dilute superfluid gas~(supersolid phase)~\cite{F1,S1, I1}. A further increase of $\epsilon_{dd}$ leads to a state where the droplets are independent and mutually incoherent, and the system does not show any extended superfluidity~(independent droplet phase) 
\footnote{Notice however that each droplet is still superfluid.}.
%



\begin{figure}[ht!]
    \centering
    \includegraphics[width=\linewidth]{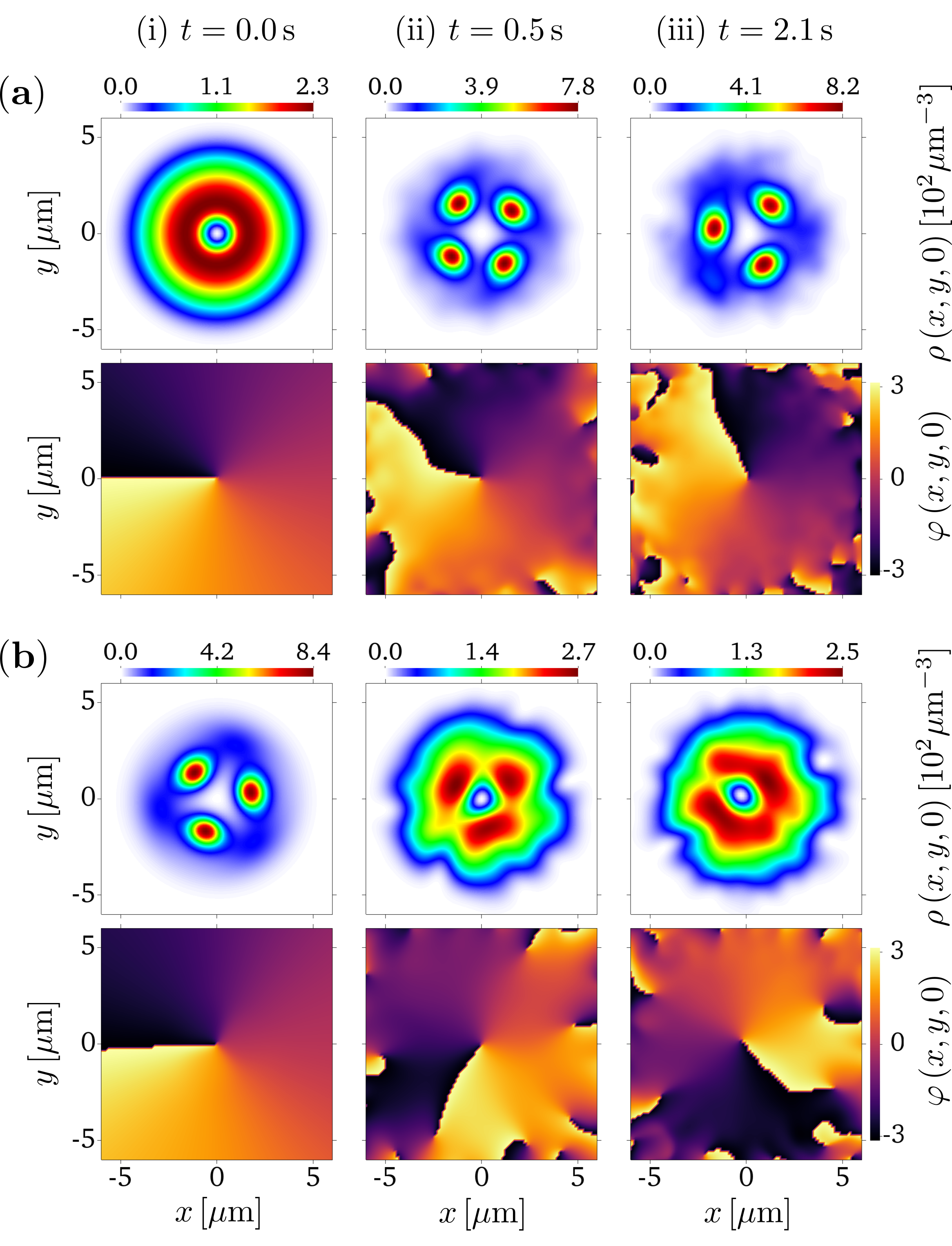}
    \caption{Density and phase profiles in the $z=0$ plane. (a)~Vortex through the superfluid-to-supersolid crossing at $t=0$, $0.5$, and $2.1\,$s. The vortex is initially created in the superfluid regime~($\Omega / \omega_\perp = 0.22$, $a=105\, a_0$), and $a$ is then linearly ramped in $100\, \mathrm{ms}$ down to $94\, a_0$, within the supersolid phase.
    (b)~Vortex through the supersolid-to-superfluid crossing for the same times. At $t=0$, the vortex is in the supersolid phase~($\Omega / \omega_\perp = 0.16$, $a=94\, a_0$). 
    Then $a$ is ramped in $100\, \mathrm{ms}$ up to  $105\, a_0$, within the superfluid regime.    
    In all cases, $\omega_\perp = 2\pi\times 60\, \mathrm{Hz}$, $\lambda = 2$, and $\varepsilon = 0$.}
    \label{fig:cross}
\end{figure}


The supersolid phase can host quantized vortices~\cite{gallemi20}.  As already anticipated in the Introduction, vortex nucleation is significantly favored by the reduced density in the inter-droplet regions, 
but vortices nestle in those interstitials, making their experimental observation much more problematic than in the superfluid phase. 
Below, we first discuss the robustness of vortices when quenching the system across the superfluid-to-supersolid transition. We then exploit 
such a robustness to design a novel protocol that first allows for a new mechanism for the nucleation of vortices in the supersolid phase and, second, for probing their existence 
by imaging them in the superfluid phase, where they are more easily detectable, also thanks to  the large increase of their core size as compared to condensates with only contact interactions.  This second step resembles the procedure used in the pioneering work of Ref.~\cite{Zwierlein05}, where the vortices created in a strongly interacting Fermi gas were imaged by quenching from the BCS to the BEC regime, where their visibility was better ensured after gas expansion. In the case of dipolar gases the procedure is more challenging because the two regimes, supersolid and superfluid, are separated by a first-order phase transition and not connected by a continuous crossover.



\paragraph{Crossing the superfluid-to-supersolid transition.--}  

We consider a BEC of $4 \times 10^4$ $^{164} \mathrm{Dy}$ atoms, confined in an axially symmetrical trap ($\varepsilon=0$) with $\omega_\perp = 2\pi\times 60\, \mathrm{Hz}$ and $\lambda = 2$. Under these conditions, the superfluid-to-supersolid transition occurs at the value $a_{\mathrm{crit}}=94.6\,a_0$~($a_0$ is the Bohr radius) corresponding to $\epsilon_{dd}=1.395$.

Ground states of the system are calculated using imaginary time evolution in the rotating frame, obtained by adding the constraint $-\Omega L_z$ to the eGPE~\eqref{eGP}, where $\Omega$ is the angular velocity, and $L_z$ the $z$ component of the angular momentum operator \footnote{The critical value $a_{\mathrm{crit}}$ increases slightly by increasing $\Omega$. Such change is however less than $0.5\%$ for the angular velocities used in this text.}. Above some critical angular velocity $\Omega_c$, vortical solutions become energetically favourable. It is important to notice that $\Omega_c$ is significantly smaller than the one required for the dynamical vortex nucleation \cite{gallemi20}, associated with a quadrupolar instability, as we discuss later.

We first consider a vortex in the superfluid phase, obtained for $a=105\,a_0 > a_\mathrm{crit}$, and $\Omega=0.22\,\omega_\perp > \Omega_c$, see Fig.~\ref{fig:cross}\,(a,\,i).
In the superfluid phase the vortex is characterized by an angular momentum $\hbar$ per particle.
Starting from this ground state configuration \footnote{All the time-dependent simulations presented in this work are performed in the laboratory reference frame.}, we ramp down in $100\,$ms the $s$-wave scattering length
to a value $a=94\,a_0<a_\mathrm{crit} $, which would correspond at equilibrium to the supersolid phase. 
 Indeed, once the transition is crossed, a strong density modulation emerges on a very short timescale, leading to the formation of droplets.  After a certain waiting time the system acquires a configuration close to the ground-state shape with a vortex in the supersolid phase~(Fig.~\ref{fig:cross}\,(a,\,iii)).
It is, however, interesting to notice that in most cases we find a transient regime~(see Fig.~\ref{fig:cross}\,(a,\,ii)) where the number of peaks is larger~(four droplets) than in the final, ground-state-like configuration~(three droplets).
 Despite the occurrence of small oscillations caused by the crossing of the first order transition, the vortex survives at the trap center, with its characteristic phase pattern. Since angular momentum is conserved during the ramping of the scattering length, and since in the supersolid the angular momentum per particle carried by the vortex is smaller than $\hbar$ due to the reduced global superfluidity~\cite{gallemi20}, the remaining angular momentum is carried by the droplets, whose centers of mass rotate in the laboratory frame with an angular velocity larger than $\Omega$.



\paragraph{Crossing the supersolid-to-superfluid transition.--}  
We carry out the same analysis in the opposite direction, following the fate of a quantized vortex initially present in the supersolid phase, an especially relevant case for the protocol discussed below.
As discussed in Ref.~\cite{gallemi20}, the angular velocity $\Omega_c$, for which the vortex becomes energetically favourable, is much smaller than the one in the superfluid phase. 
In Fig.~\ref{fig:cross}\,(b,\,i), we consider a configuration with $a=94\,a_0$ and $\Omega=0.16\,\omega_\perp$, slightly higher than the critical value  $\Omega_c$.
The created vortex is characterized by an angular momentum per particle of $0.87\, \hbar$. After ramping in $100\, \mathrm{ms}$ the scattering length 
up to $a=105\, a_0$  to reach the superfluid phase, we find that the vortex remains clearly visible~(Figs.~\ref{fig:cross}\,(b,\,ii) and~(b,\,iii)). Note, however, that the density profile preserves some density modulations, which are the residue of the original droplets characterizing the supersolid phase.
%
Moreover, since the overall angular momentum must be preserved, the larger angular momentum associated with the vortex in the superfluid phase ($\hbar$) is compensated by 
the rotational motion of the density modulations, and by the occurrence of anti-vortices located near the border of the atomic cloud, as well as, in some cases, by a slight displacement of the vortex core from the center of the trap.



\begin{figure}[t!]
    \centering
    \includegraphics[width=\linewidth]{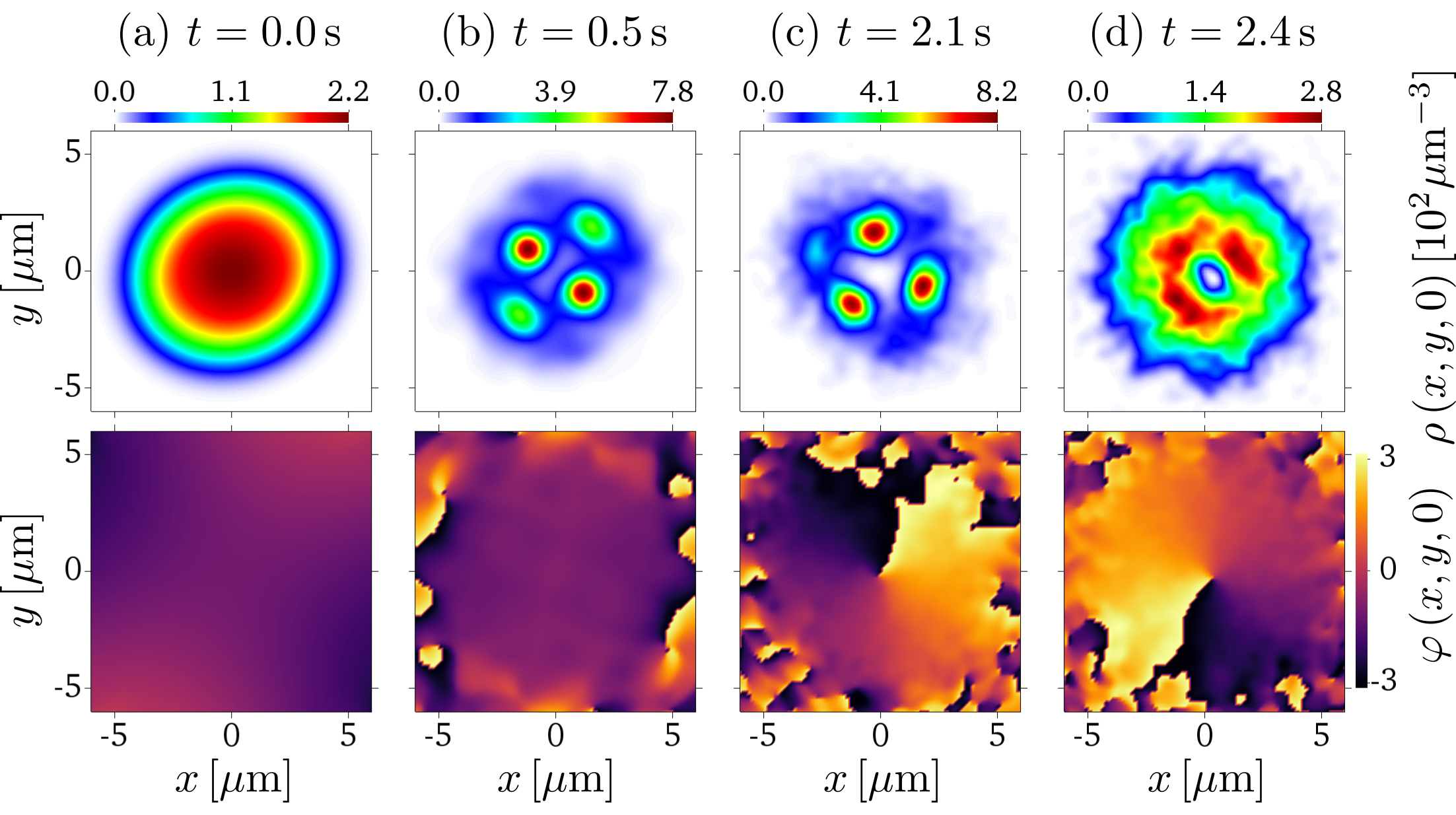}
    \caption{
   Density and phase profiles in the $z=0$ plane, showing vortex nucleation employing the protocol discussed in the text. 
    (a) Initial vortex-free superfluid with a scattering length $a = 105 \, a_0$, confined in a slightly deformed harmonic trap ($\omega_\perp = 2\pi\times 60\,\mathrm{Hz}$, $\lambda = 2$, $\varepsilon = 6.6\%$), rotating with an angular velocity $\Omega= 0.3\,\omega_\perp$.
    (b) The scattering length is linearly ramped in $100\,\mathrm{ms}$ down to $a_\mathrm{s} = 94 \, a_0$ resulting in a transition to the supersolid phase.
    (c) After some time a vortex is nucleated at the center of the trap.
    (d) The isotropy of the trap is restored ($\varepsilon = 0$) and the scattering length is linearly ramped in $100\,\mathrm{ms}$ up to the initial value, resulting in a superfluid with a readily detectable vortex core.}
    \label{fig:protocol}
\end{figure}




\paragraph{Protocol for vortex nucleation and detection.--} 
We are now ready to discuss  our protocol which combines the favorable nucleation mechanism of quantized vortices exhibited by the supersolid phase with their topological robustness  when the supersolid-superfluid phase transition is crossed. Our starting point is a slowly rotating trapped dipolar gas in the superfluid phase ($a = 105\,a_0$), obtained by a sudden introduction of rotation to the superfluid ground state in a slightly deformed trap in the $xy$-plane, and letting it equilibrate for $200\,\mathrm{ms}$ (see Fig.~\ref{fig:protocol}\,(a)).  In the laboratory frame this corresponds to choosing a harmonic potential of the form
\begin{align}
V_{ext}(t) &= \frac{m}{2}\omega_\perp^2\left[ (1-\varepsilon)(x\cos(\Omega t)+y\sin(\Omega t))^2\right\delimiter 0 \nonumber\\
&+\left\delimiter 0 (1+\varepsilon) (-x\sin(\Omega t)+y\cos(\Omega t) )^2+ \lambda^2z^2 \right] 
\label{Vho}    
\end{align}

We choose a slightly deformed trap ($\varepsilon = 6.6\%$) and an angular velocity ($\Omega =$ $0.3\,\omega_\perp$) such that the system is unable to nucleate vortices in the superfluid phase, as the quadrupole  dynamical instability occurs at $0.45\,\omega_\perp$~\footnote{This value remains almost constant in the whole superfluid region.}. 
The parameters are  instead large enough for vortex nucleation once the system enters the supersolid phase. Therefore we reduce the value of the scattering length with a linear ramp in $100\, \mathrm{ms}$ down to 
$a=94\,a_0$.  After entering the supersolid phase first droplets are formed~(Fig.~\ref{fig:protocol}\,(b)) and, after a while, a vortex is nucleated in the center~(Fig.~\ref{fig:protocol}\,(c)). Notice that the time scale for this process is slow in the present simulation. We expect, however, that in a real experimental situation the time scale will be much faster, as a consequence of thermal noise, which is not accounted for in our calculations. When the vortex is formed~(Fig.~\ref{fig:protocol}\,(c)) we restore the isotropy of the trap ($\varepsilon=0$) in order to ensure the robustness of the topological configuration associated with the vortex and the conservation of angular momentum. We ramp the scattering length back to its initial value (following a similar ramp) and after a while~(Fig.~\ref{fig:protocol}\,(d)) the system enters again the superfluid phase. We then recover a very similar configuration as 
that of Fig.~\ref{fig:cross}\,(b,\,iii).



\begin{figure}[t!]
    \centering
    \includegraphics[width=\columnwidth]{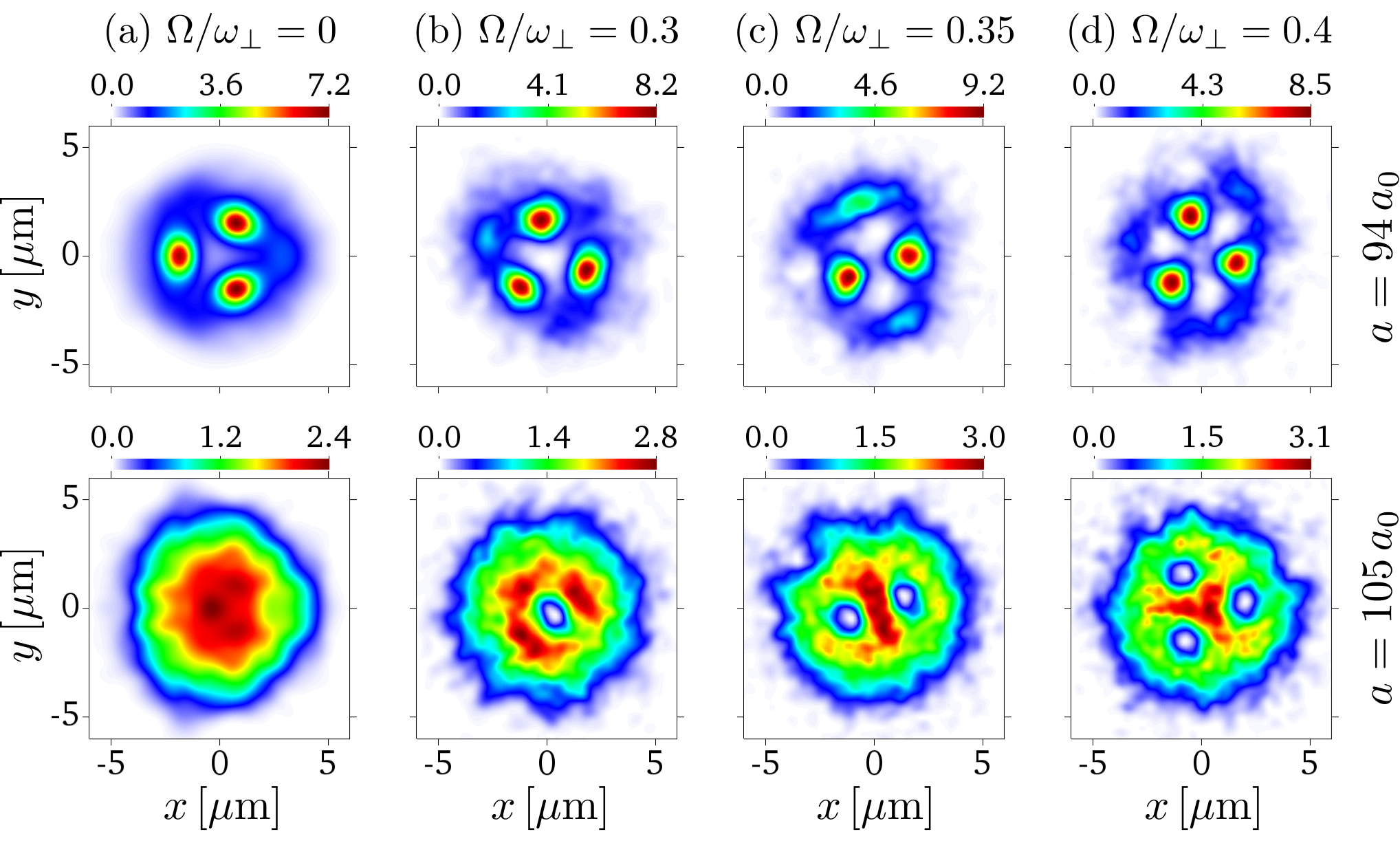}
    \caption{ Density profiles in the $z=0$ plane. Results for the dynamical protocol for different angular velocities: $\Omega/\omega_\perp=0$~(a), $0.3$~(b), $0.35$~(c), and $0.4$~(d). 
    The parameters and procedure are the same as in Fig.~\ref{fig:protocol}. 
    The upper row corresponds to the configurations in the supersolid phase at $t=2.1\,\mathrm{s}$, before inverting the ramp of the scattering length. 
    The lower row corresponds to the final configuration in the superfluid phase (at $t=2.4\,\mathrm{s}$) after ramping back the scattering length. 
    Note that imaging in the superfluid phase should easily reveal the presence of no vortex, one, two and three vortices, respectively.
    The final angular momentum per particle (once the isotropy of the $xy$ trapping is restored) is (a) $L = 0$, (b) $L = 1.03\,\hbar$,  (c) $L = 1.85\,\hbar$, and (d) $L = 2.42\,\hbar$.}
    \label{fig:3}
\end{figure}


The same protocol may be employed for the nucleation of more than one vortex when increasing the angular velocity $\Omega$ of the rotating trap. In Fig.~\ref{fig:3}, we show 
our results for different angular velocities in our protocol. The upper panel shows the atomic cloud in the supersolid phase right before inverting the ramping of the scattering length. The lower panel depicts the final density distribution after ramping back the scattering length. 
The case with $\Omega=0$ is important, since it clearly shows that despite the strong density modulation in the supersolid regime, once moving back into the superfluid 
no core appears, the final density remains smooth and characterized by a maximum in the center, very similar to the initial equilibrium configuration. 
By increasing $\Omega$ we eventually observe one vortex nucleated in the center using $\Omega=0.3\,\omega_\perp$ (same  as Figs.~\ref{fig:protocol}\,(c) and (d)), two vortices using 
$\Omega= 0.35\,\omega_\perp$ and three vortices using $\Omega =0.4\,\omega_\perp$.   Note that in all cases 
the vortices are nucleated in the supersolid phase, since  the angular velocity is not large enough to create vortices in the superfluid.



\paragraph{Conclusions.--} 
We have studied vortices in a dipolar condensate when crossing the superfluid-to-supersolid transition. 
We have proposed in particular a novel protocol that should permit under realistic conditions to nucleate and detect quantized vortices in a dipolar supersolid, a major 
hallmark of superfluidity. The method is based on a controlled ramp of the scattering length across the superfluid-to-supersolid transition, employing the 
very nature of the supersolid to induce vortex nucleation. Although vortex detection is difficult in the supersolid since vortices gather in regions of 
very low density, a ramp back into the superfluid  permits an easy imaging of the vortex core, even more so than in contact-interacting 
condensates due to the significantly larger vortex size in a dipolar BEC. Very recently quantized vortices have been actually observed in the superfluid phase of a dipolar gas \cite{VortexIBK}.



\acknowledgements 
We thank A. Gallem\'i for interesting discussions. This work was supported by Q@TN (the joint lab between University of Trento, FBK - Fondazione Bruno Kessler, INFN - National Institute for Nuclear Physics and CNR - National Research Council) and  the Provincia Autonoma di Trento. L.S. acknowledges support of the Deutsche Forschungsgemeinschaft (DFG, German Research Foundation) under Germany's Excellence Strategy -- EXC-2123 QuantumFrontiers -- 390837967, and FOR 2247. S.M.R. acknowledges support from the Alexander von Humboldt Foundation. A. R. acknowledges support of the Italian MUR under the PRIN2017 project CEnTraL (ProtocolNumber 20172H2SC4). M.S. acknowledges funding provided by the Institute of Physics Belgrade, through the grant by the Ministry of Education, Science, and Technological Development of the Republic of Serbia. We acknowledge the CINECA award under the ISCRA initiative, for the availability of high performance computing resources and support.

\bibliography{biblio-inertia.bib}

\end{document}